\documentclass{article}

\usepackage{arxiv}

\usepackage[utf8]{inputenc} 
\usepackage[T1]{fontenc}    
\usepackage{hyperref}       
\usepackage{url}            
\usepackage{booktabs}       
\usepackage{amsfonts}       
\usepackage{nicefrac}       
\usepackage{microtype}      
\usepackage{lipsum}		
\usepackage{graphicx}
\usepackage{cite}
\usepackage{doi}

\usepackage{amsmath,amssymb,amsfonts}
\usepackage{subcaption}
\usepackage{here}

\title{Can AI with High Reasoning Ability Replicate Human-like Decision Making in Economic Experiments?}

\author{ \href{https://orcid.org/0000-0002-4774-1506}{\includegraphics[scale=0.06]{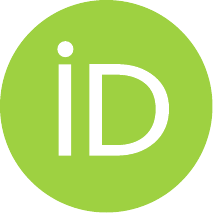}\hspace{1mm}Ayato~Kitadai}\thanks{\texttt{a.kitadai@css.t.u-tokyo.ac.jp}} \\
	School of Engineering\\
	The University of Tokyo\\
	\And
	\href{https://orcid.org/0000-0002-6863-5139}{\includegraphics[scale=0.06]{orcid.pdf}\hspace{1mm}Sinndy~Dayana~Rico~Lugo} \\
        School of Engineering\\
	The University of Tokyo\\
        \And
        Yudai~Tsurusaki \\
        School of Engineering\\
	The University of Tokyo\\
        \AND
        Yusuke~Fukasawa\\
        School of Engineering\\
	The University of Tokyo\\
        \And
	\href{https://orcid.org/0000-0002-6411-8716}{\includegraphics[scale=0.06]{orcid.pdf}\hspace{1mm}Nariaki~Nishino} \\
        School of Engineering\\
	The University of Tokyo\\
}

\renewcommand{\shorttitle}{Can AI with High Reasoning Ability Replicate Human-like DM in Economic Experiments?}

\hypersetup{
pdftitle={Can AI with High Reasoning Ability Replicate Human-like Decision Making in Economic Experiments?},
pdfsubject={q-bio.NC, q-bio.QM},
pdfauthor={Ayato~Kitadai, Nariaki~Nishino},
pdfkeywords={large language models, economic experiment, multi agent simulation, ultimatum game, decision making},
}

\begin{document}
\maketitle

\begin{abstract}
Economic experiments offer a controlled setting for researchers to observe human decision-making and test diverse theories and hypotheses; however, substantial costs and efforts are incurred to gather many individuals as experimental participants. To address this, with the development of large language models (LLMs), some researchers have recently attempted to develop simulated economic experiments using LLMs-driven agents, called generative agents. If generative agents can replicate human-like decision-making in economic experiments, the cost problem of economic experiments can be alleviated. However, such a simulation framework has not been yet established. Considering the previous research and the current evolutionary stage of LLMs, this study focuses on the reasoning ability of generative agents as a key factor toward establishing a framework for such a new methodology. A multi-agent simulation, designed to improve the reasoning ability of generative agents through prompting methods, was developed to reproduce the result of an actual economic experiment on the ultimatum game. The results demonstrated that the higher the reasoning ability of the agents, the closer the results were to the theoretical solution than to the real experimental result. The results also suggest that setting the personas of the generative agents may be important for reproducing the results of real economic experiments. These findings are valuable for the future definition of a framework for replacing human participants with generative agents in economic experiments when LLMs are further developed.
\end{abstract}

\section{Introduction}
Economic experiments are a relevant approach in economics to provide researchers with a controlled environment to observe and analyze human behavior in response to various economic conditions to test theories and hypotheses, assess the impact of policy changes, and understand decision-making processes. However, it is necessary to gather many individuals as subjects to get sufficient experimental data that can be judged as valuable and reproducible \cite{Latuszynska-2015} for such testing, commonly incurring huge costs and efforts to provide the subjects with enough motivation to participate and provide serious responses \cite{Kunze-2006}. 

Considering this problem, a novel simulation approach called ``Large Language Models-driven multi-agent simulation (LLMs-driven MAS)'' has recently received the attention of some researchers in economics and behavioral fields. In LLMs-driven MAS, each agent, called a generative agent, makes decisions using LLMs, making it different from conventional MAS where agents follow completely predefined rules. Especially, the most recent LLMs-driven MAS that can generate human-like responses using natural languages are becoming more attractive \cite{Park-2023}.

Particularly, some researchers have recently worked on utilizing LLMs-driven MAS in economic experiments, but they are still under challenges. On the one hand, some studies succeeded in observing the same tendency of well-known classical results of some economic experiments in LLMs-driven MAS \cite{Aher-2023, Horton-2023}. On the other hand, some articles report that LLMs-driven MAS had shown both consistent and inconsistent data with real economic experiments \cite{Brookins-2023, Tsuchihashi-2023, Guo-2023, Phelps-2023}. In addition, some researchers affirm that the result of LLMs-driven MAS for economic experiments depends on the setting of the simulation such as the persona of generative agents and the instructions to agents as subjects \cite{Horton-2023, Phelps-2023, Han-2023}. Thus, it can be said that the way to apply LLMs-drive MAS into economic experiments has not been established and further study is necessary on this topic.

As a first step to establishing a framework for such a new methodology, in Kitadai et al. (2023), we conducted LLMs-driven MAS to explore the reproducibility of the outcome of the economic experiment on the one-shot ultimatum game published by Lin et al. (2020). The ultimatum game is a well-known classical theme of economic experiments which suggests the gap between theoretical equilibrium and actual human behavior\cite{Guth-1982, Lin-2020}. In the study, various settings of prompt and parameter were used to analyze a general proper configuration of LLMs-driven MAS for economic experiments. As a result, we found a suitable setting to reproduce the distribution of the original data for the proposer side, but could not for the responder side\cite{Kitadai-2023}.

Based on that, this paper presents the next step of our research toward establishing the general way of applying LLMs-driven MAS into economic experiments. As we explain in detail later, it was considered that the reasoning ability of the generative agents in our previous simulation was insufficient to reproduce the original result for the responder side. Therefore, this research aims to clarify if LLMs-driven MAS can reproduce the result of economic experiments by improving the reasoning ability of generative agents. In particular, we conducted LLMs-driven MAS, using an updated version of the framework based on Kitadai et al. (2023) and compared the results of simulations through various methods to improve the reasoning ability of agents to reproduce the outcome of the one-shot ultimatum game shown in Lin et al. (2020).

The contribution of this research will be valuable for utilizing LLMs-driven MAS in economic experiments even in the future when LLMs with higher reasoning abilities are in widespread use. With the rapid development of the computer science field, models of LLMs with higher reasoning abilities are appearing and are going to prevail. Considering this trend, it can be said that taking the level of reasoning ability of LLMs into account is important to establish the way to apply LLMs-driven MAS into economic experiments.

The reminder of this paper consists of the theoretical background presented in Section \ref{sec:theoretical}, the research methodology in Section \ref{sec:methodology}, the results shown in Section \ref{sec:results}, and the discussion and conclusions presented in Section \ref{sec:discussion}.

\section{Theoretical Background}\label{sec:theoretical}
\subsection{Ultimatum Game}
The ultimatum game involves a sequential decision-making process in which two players determine the distribution of a divisible good. Although there are both one-shot and repeated variations of this game, our investigation specifically concentrated on the one-shot game ultimatum game. It comprises the following two steps \cite{Kitadai-2023}:

\begin{enumerate}
    \item The player making the offer suggests a partition of divisible the good. As an example, with a total of $100$ coins in play, the offering player may propose taking $70$ coins for himself, leaving the responding player with $30$ coins.
    \item The player responding has the choice to either accept or reject the proposed division. Upon acceptance, the suggested distribution is implemented; however, in the case of rejection, both players receive nothing.
\end{enumerate}

In a mathematical form, the divisible good is defined as $R\in\mathbb{R}$. The strategies of players are: $1$ is $s_1\in [0, 1]$ for player $1$ and $s_2\in \{\rm{accept}, \rm{reject}\}$ for player $2$. The payoffs of players are:
\begin{equation*}
    (u_1, u_2) = 
    \left\{
    \begin{alignedat}{2}
        &(R(1-s_1), Rs_1)\quad &&\text{if $s_2 = \rm{accept}$},\\
        &(0, 0) &&\text{if $s_2 = \rm{reject}$}.
    \end{alignedat}
    \right.
\end{equation*}
where $u_i$ is the payoff of player $i$.

Although $s_2 = \rm{accept}$ is the weakly dominant strategy of player $2$, in economic experiments, it is common to observe decisions of $s_2 = \rm{reject}$ when the value of $s_1$ is low \cite{Guth-1982}. A commonly proposed explanation for this phenomenon is that responders may choose to sacrifice their payoffs to resist perceived unfairness, particularly when their potential earnings are lower than those of the proposer.

To analyze economic experiments conducted globally during classes, Lin et al. (2020) employed data from MobLab, a well-known educational platform for such experiments. Specifically, the study focused on double auction and ultimatum games, revealing consistent insights across different cultures and regions\cite{Lin-2020}. Our current investigation concentrates on two outcomes derived from the one-shot ultimatum game, which are different from the theoretical solution, subgame perfect equilibrium, of this game: $(s_1, s_2) = (0, \rm{accept})$.

\begin{figure}[htbp]
    \centering
    \begin{subfigure}[t]{0.45\textwidth}
        \centering
	\includegraphics[width=\textwidth]{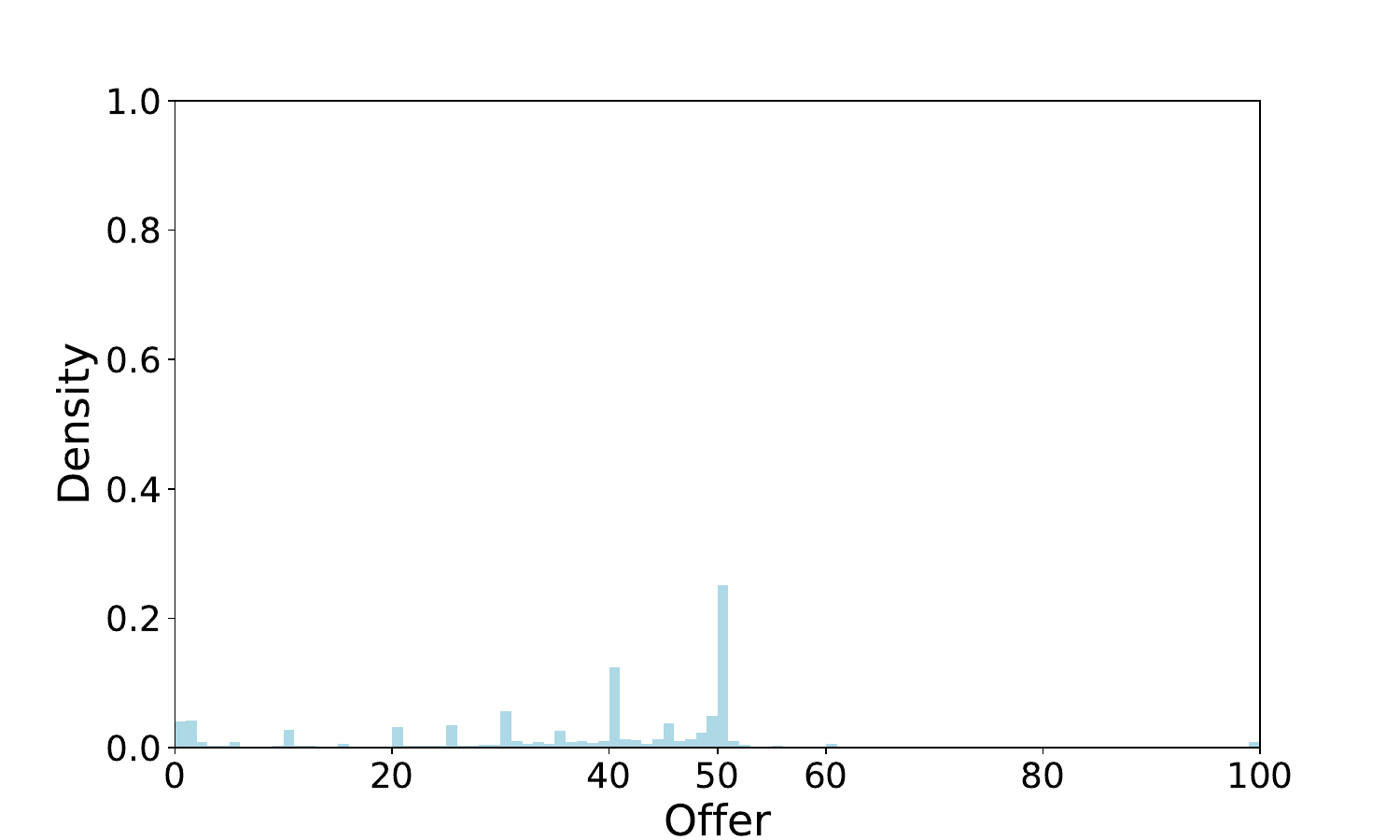}
        \subcaption{Proposer side}\label{fig:proposer-raw}
    \end{subfigure}
    \begin{subfigure}[t]{0.45\textwidth}
	\centering
        \includegraphics[width=\textwidth]{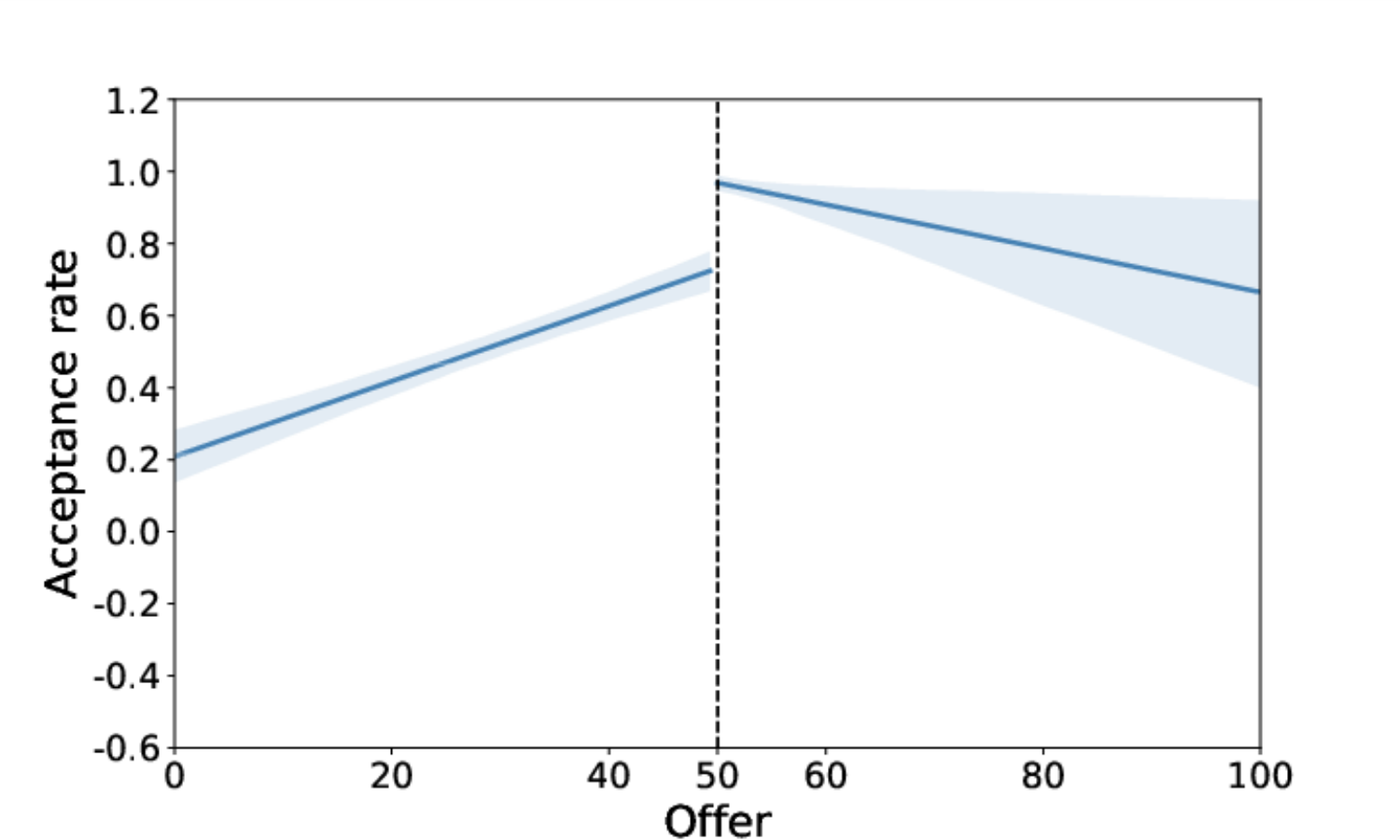}
        \subcaption{Responder side}\label{fig:responder-raw}
    \end{subfigure}
    \centering
    \caption{Results of 1000 sample data of economic experiments for one-shot ultimatum game in Lin et al. (2020)}
\end{figure}

Based on a sample of $1000$ experimental data for the proposer side published by Lin et al. (2020), we plot Figure~\ref{fig:proposer-raw}. The horizontal axis shows the proposer values and the vertical axis displays their normalized frequency. The distribution peaks at a $50$ offer, with few proposals beyond.

Similarly, in Figure~\ref{fig:responder-raw}, the horizontal axis represents the proposer values, while the vertical axis indicates the corresponding acceptance rate. The straight line represents the outcomes of piecewise linear regression, while the shaded area indicates the confidence interval. A notable abrupt increase is evident at the $50$ offer. Moreover, there is a discernible upward trend in the acceptance rate up to $50$ as the offer value increases.

\subsection{Generative Pretrained Transformer}\label{subsec:gpt}
Generative Pretrained Transformer (GPT) is a type of LLM developed by OpenAI\textregistered, which is the most famous and prevailing type at the moment of this research. After the announcement of GPT-3, a type of GPT, by Brown et al. (2020), the dizzying development of the Computer Science field was accelerated and is still rapidly developing~\cite{Brown-2020}. Even today, GPT is a standard of LLMs and widely used both in research and business fields and we also used a model of GPT in our research.

There are various types of GPT. Kitadai et al. (2023) used ``gpt-3.5-turbo-0613'' which is the version of GPT-3.5 as of June 23, 2023. Although GPT-4 is known as a higher performance model of GPT-3.5, because ``gpt-4-0613'' was more time-consuming to make a response, it was not used in Kitadai et al. (2023).
However, ``gpt-4-1106-preview'', which is the version of GPT-4 as of November 11, 2023, responds quicker for a large sample of simulation and has higher reasoning ability than GPT-3.5. 

Not only the models but diverse prompting techniques are also being developed to increase the reasoning ability of LLMs. Because of the large number of such methods currently available, a selection was necessary. Specifically, the following two prompting methods were chosen in this study, because they are widely known and established as valid methodologies:
\begin{itemize}
    \item Few-shot prompting\cite{Brown-2020}:\\
    Putting some examples of outputs into the prompt.


    \item Chain of Thoughts (CoT)\cite{Wei-2022}:\\
    Putting some examples of outputs with their corresponding reason into the prompt and making agents respond in the same way.
\end{itemize}

In comparison to Few-shot prompting, usual prompts just putting simple instructions without any examples are called Zero-shot prompting. To analyze the improvement in reasoning ability, the prompting methods used in this study are, in the reasoning ability order: CoT, Few-shot prompting, and Zero-shot prompting.

\subsection{Literature Review}
Many researchers are studying the application of generative agents in social science, including economics~\cite{Li-2023, Brand-2023, Bybee-2023, Lore-2023, Akata-2023} and psychology~\cite{Dillion-2023, Hagendorff-2023, Harding-2023, Argyle-2023, Binz-2023}. In particular, Aher et al. (2022) utilized generative agents as subjects in economic experiments at first. They proposed ``Turing Experiments (TEs)'' as an evaluating method to measure how much LLMs can simulate various human behavioral aspects. TEs consist of four classic social scientific experiments that are not only economic experiments but also include the ultimatum game's responder side~\cite{Aher-2023}. 

After that, some researchers reported that LLMs-driven MAS can reproduce the results of real economic experiments. The most representative one is Horton (2023), who picked up three classical themes related to the preferences of people and researched the effect of the presence of a minimum wage~\cite{Horton-2023}.

In contrast, other academics argue that caution should be taken when replacing human subjects with generative agents in such a type of experiment. Brookins et al. (2023) made generative agents play the dictator game and the prisoner's dilemma. As a result, in both games, generative agents showed fair decision making more frequently than human subjects~\cite{Brookins-2023}. Guo (2023) focused on the finitely repeated ultimatum and prisoner's dilemma games, finding that generative agents can partly exhibit behavior consistent with human behavior by using well-crafted prompts~\cite{Guo-2023}.

Phelps and Russell (2023) also simulated the finitely repeated prisoner's dilemma. They used four kinds of bots as the opponent player of the generative agents: always cooperating bot, always defecting bot, and bots who cooperate/defect in the first round and after that imitate the choice of the generative agent in the previous round. As a result, generative agents could not appropriately behave in response to different strategies of the opponent~\cite{Phelps-2023}. Tsuchihashi (2023) simulated the first price auction (FPA) and second price auction (SPA) using generative agents. Although it is known that human subjects tend to bid higher than their willingness to pay in both FPA and SPA, generative agents did not show such trends~\cite{Tsuchihashi-2023}.

Considering these, it can be said that how to apply LLMs-drive MAS into economic experiments has not been established yet and further study is necessary on this topic. Towards the establishment of such a methodology to reproduce the result of economic experiments, Kitadai et al. (2023) made the first contribution. This research is the next step of Kitadai et al. (2023) and explores the influence of the reasoning ability of generative agents to replicate the real economic experiments.

\subsection{Our Previous Study}
In our previous study, Kitadai et al. (2023), we searched for reproducing the result of the ultimatum game of Lin et al. (2020) by using GPT-driven MAS focusing on the effects of the following three aspects:
\begin{itemize}
    \item How the amount of money subjects will receive at the end of the experiment is determined,
    \item Whether explicit instructions aimed at maximizing rewards are given, 
    \item The setting of the parameter ``temperature'' in GPT.
\end{itemize}

Herein, ``temperature'' is a parameter in GPT that influences the randomness and creativity of its output, ranging from $0$ to $2$. Lower temperatures lead to more predictable, conservative outputs by favoring higher probability words, and higher temperatures increase creativity and variability, encouraging the model to choose less likely words, which can result in more novel and diverse responses but also raises the likelihood of irrelevant or nonsensical content.

\begin{figure}[htbp]
    \centering
    \begin{subfigure}[t]{0.45\textwidth}
        \centering
	\includegraphics[width=\textwidth]{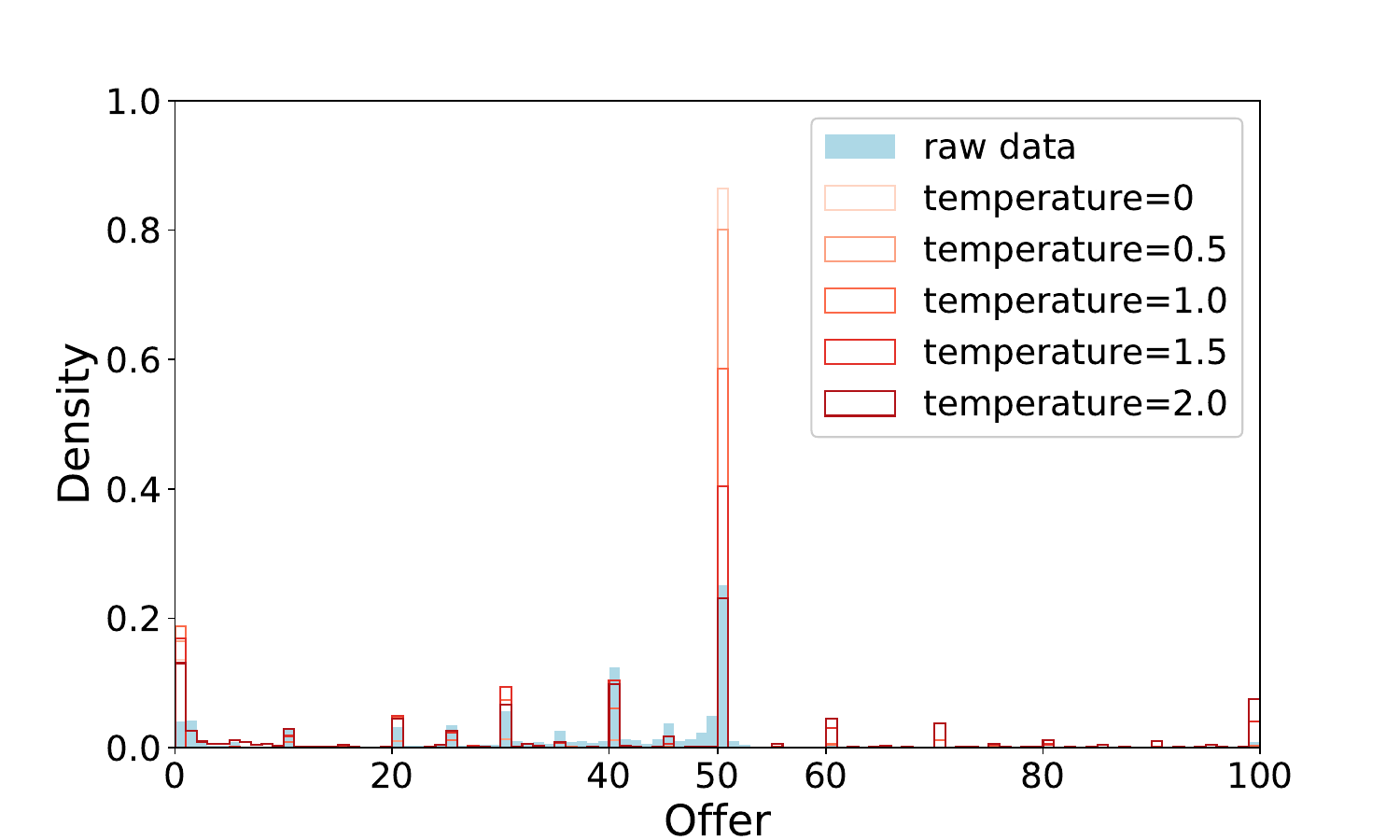}
        \subcaption{Proposer side}\label{fig:proposer-pre-result}
    \end{subfigure}
    \begin{subfigure}[t]{0.45\textwidth}
	\centering
        \includegraphics[width=\textwidth]{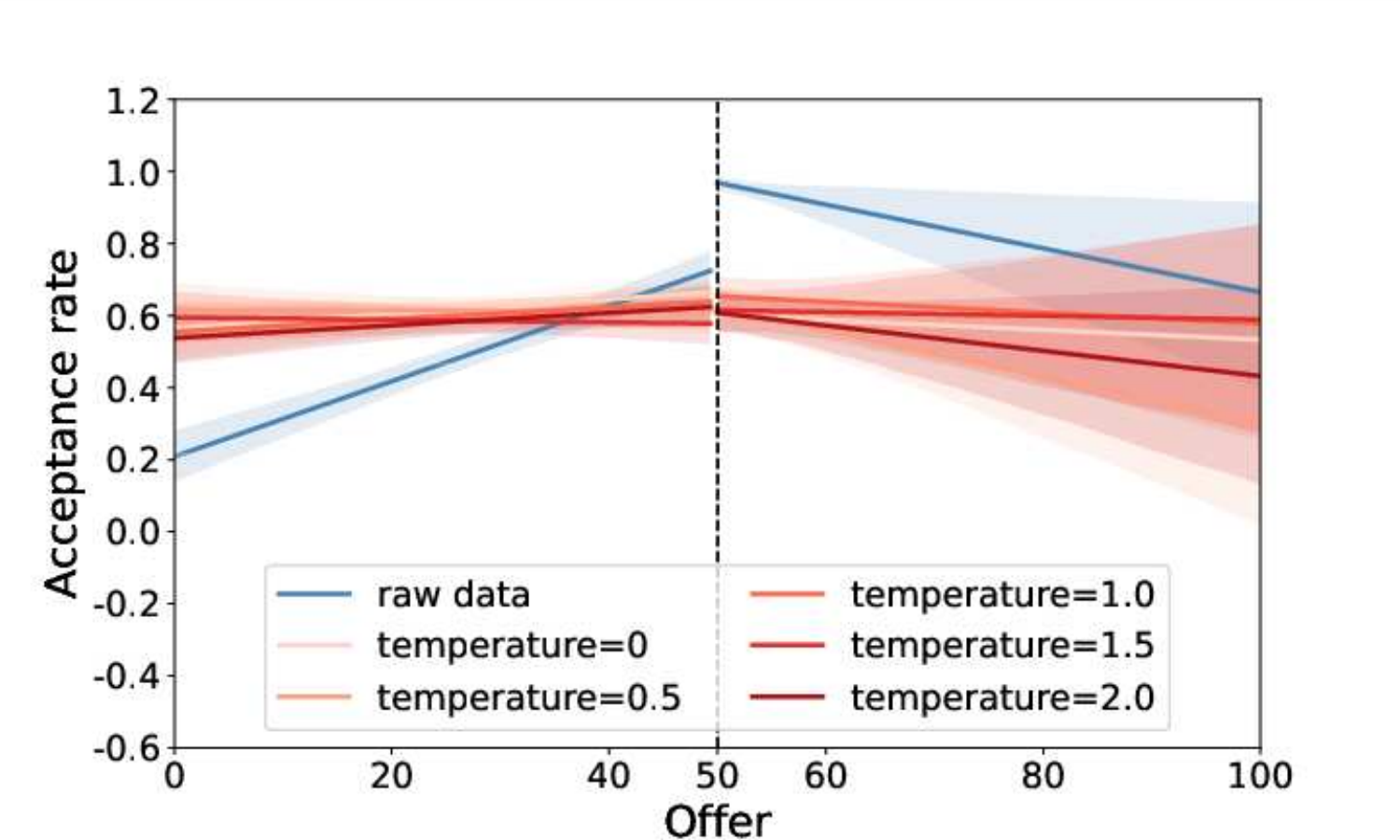}
        \subcaption{Responder side}\label{fig:responder-pre-result}
    \end{subfigure}
    \centering
    \caption{A result of simulations in Kitadai et al. (2023)}\label{fig:result-kitadai}
\end{figure}

A set of $1000$ simulations was done based on $4\times 2\times 5 = 40$ variations in each of the points listed above. As a result, for the proposer side, it was found that not giving explicit instructions to maximize profits, specifying the amount to be received by the participants at the end of the experiment, and a higher value of ``temperature'' made the simulation result closer to the actual one. At the same time, however, any simulation settings reproduced the actual result for the responder side. Figure~\ref{fig:result-kitadai} shows the result obtained when no instruction to maximize profits was given, telling that each coin obtained in the game would be redeemed $100$ dollars~\cite{Kitadai-2023}. In the figure, the data from MobLab used by Lin et al. (2020) is plotted in blue, while the simulation results for the five temperature settings, $0, 0.5, 1.0, 1.5, 2.0$, are depicted in different colors.

Figure~\ref{fig:proposer-pre-result} presents the result for the proposer side, where the horizontal axis corresponds to the offer value and the vertical axis is the density. A peak at $50$ value of the offer appeared, being consistent with the actual experimental result. Nevertheless, Figure~\ref{fig:responder-pre-result} shows the result for the responder side, where the offer value is on the horizontal axis and the acceptance rate is on the vertical one. In this case, the responses of generative agents are almost not sensitive to variations in the proposed values. In particular, low acceptance rates against offers greater than half are not expected and tend to be irrational. Therefore, we considered that the reasoning ability of agents was not enough high to reach the consistency with the actual economic experiment.

Regarding the setting of persona, Kitadai et al. (2023) generated the information of gender, age, and personality at random by using ``gpt-3.5-turbo-0613''. However, such random generation of persona can be noise to reveal the effect of the reasoning ability of generative agents because the result of LLMs-driven MAS for economic experiments depends on the persona configuration \cite{Horton-2023, Phelps-2023, Han-2023}. Subsequently, the step of the persona setting for each agent in the simulation of the current research was eliminated.

\section{Methodology}\label{sec:methodology}
The prompt of simulations conducted in this study followed the framework below which is the updated version of Kitadai et al. (2023) as mentioned in the previous section.

\begin{itemize}
    \item[1.] Explanation of the ultimatum game
    \item[2.] Description of the situation in which the agent makes a decision and instructions for decision-making. Each prompting method (Few-shot, CoT, or Zero-shot) was introduced here
    \item[3.] Specification of the output format 
\end{itemize}

All prompts used in this study were written in English. This choice was made due to known reductions in LLM performance when prompts are written in languages other than English \cite{etxaniz-2023}. Moreover, the descriptions for the ultimatum game and the scenario for decision-making were the same as in our previous work. 

Regarding the second item in the framework, based on the insights obtained in Kitadai et al. (2023), in the prompt, agents were tasked with deciding on the distribution of a non-real-world currency of $100$ coins, and it was explained that each coin will be redeemed to $100$ dollars at the end of the game, but the instruction to maximize their profit was not mentioned.

Using this structure of prompt, we conducted the four patterns of simulations described in Table~\ref{tb:sim-settings}. Regarding the GPT model in the pattern A, the same combination of settings with Kitadai et al. (2023), we introduced the older one, gpt-3.5-turbo-0613, as a benchmark to compare the reasoning ability of generative agents. Regarding the temperature parameter, we considered five variations: $0, 0.5, 1.0, 1.5, 2.0$ as Kitadai et al. (2023) did for patterns A, B, and C. However, in pattern D, we did not consider the value $2.0$. This is because, even with the latest model, only one response per hour was obtained by using CoT with temperature $2.0$, becoming an excessively time-consuming simulation setting to be an alternative to economic experiments.

\setlength{\tabcolsep}{10pt}
\begin{table}[H]
\centering
\caption{Four Patterns of Simulation Settings}\label{tb:sim-settings}
\begin{tabular}{c||c|c|c}
\hline
Pattern & GPT Model & Prompt Method & Temperature Range \\ \hline\hline
A & gpt-3.5-turbo-0613 & Zero-shot & $0, 0.5, 1.0, 1.5, 2.0$ \\ \hline
B & gpt-4-1106-preview & Zero-shot & $0, 0.5, 1.0, 1.5, 2.0$ \\ \hline
C & gpt-4-1106-preview & Few-shot & $0, 0.5, 1.0, 1.5, 2.0$ \\ \hline
D & gpt-4-1106-preview & CoT & $0, 0.5, 1.0, 1.5$ \\ \hline
\end{tabular}
\end{table}

To use Few-shot and CoT promptings, some examples put in prompt needed to be prepared. As for few-shot prompting, we used $10$ data picked up from raw data in Lin et al. (2020). As for CoT, we respectively made a reason and added it to each sample data used in few-shot, and put them into the prompt. The prompts used in this research are put in the Appendix. For each combination of prompting method, temperature, and proposer or responder side, $1000$ agents were generated and made decisions independently.

\section{Results}\label{sec:results}

\subsection{Proposer Side}
\begin{figure}[htbp]
    \centering
    \begin{subfigure}[t]{0.45\textwidth}
        \centering
	\includegraphics[width=\textwidth]{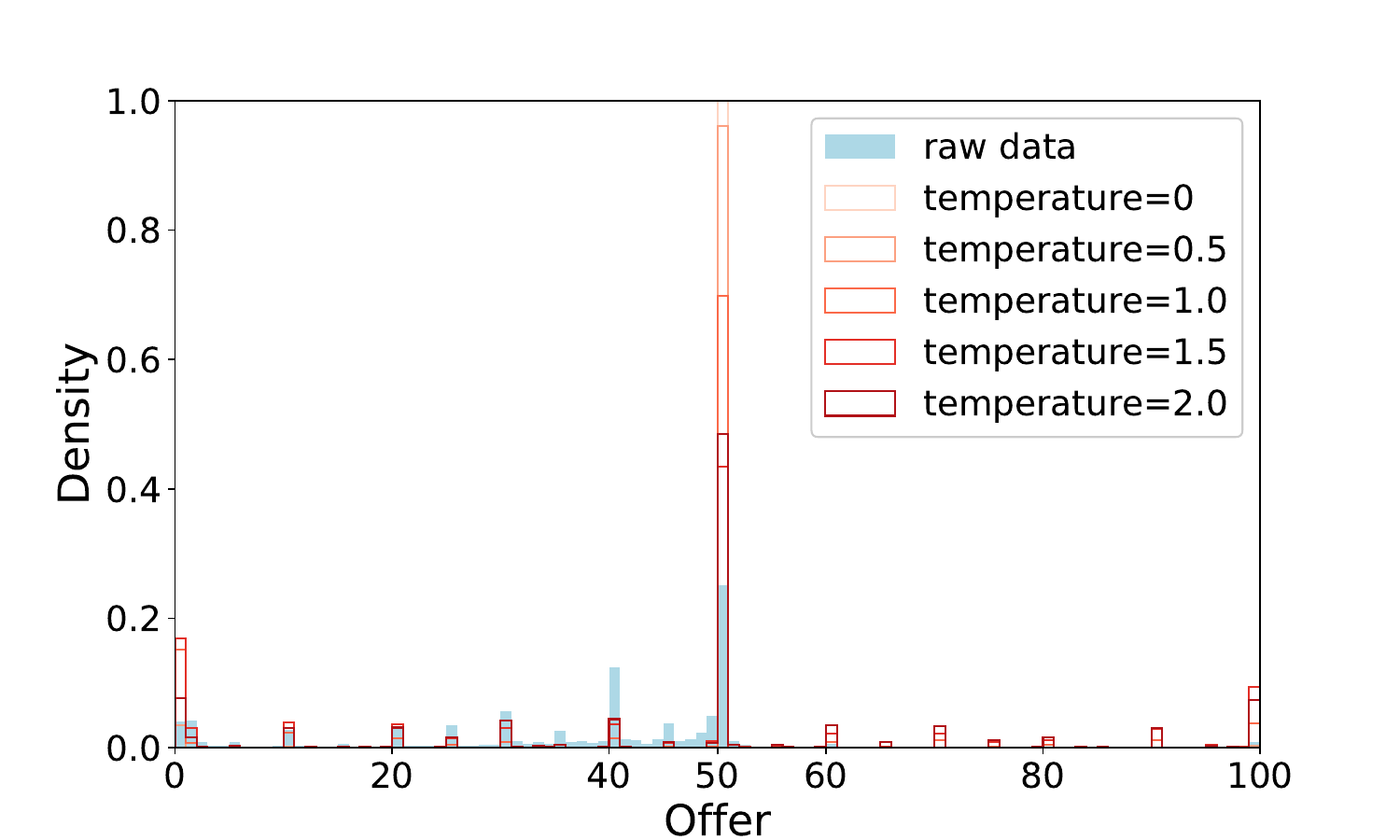}
        \subcaption{Pattern A}\label{subfig:pro-zero-3.5}
    \end{subfigure}
    \begin{subfigure}[t]{0.45\textwidth}
	\centering
        \includegraphics[width=\textwidth]{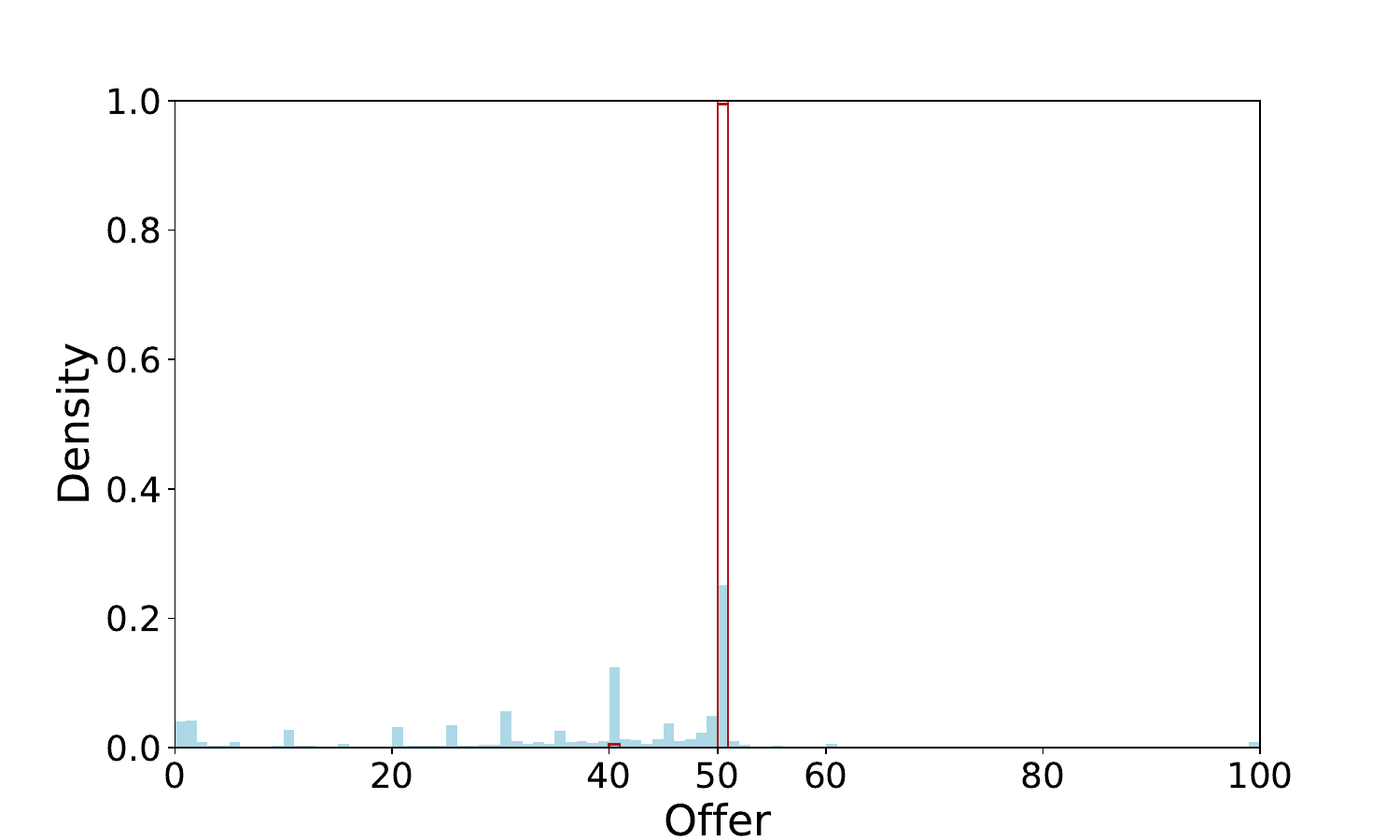}
        \subcaption{Pattern B}\label{subfig:pro-zero-4}
    \end{subfigure}
    \begin{subfigure}[t]{0.45\textwidth}    
        \centering
        \includegraphics[width=\textwidth]{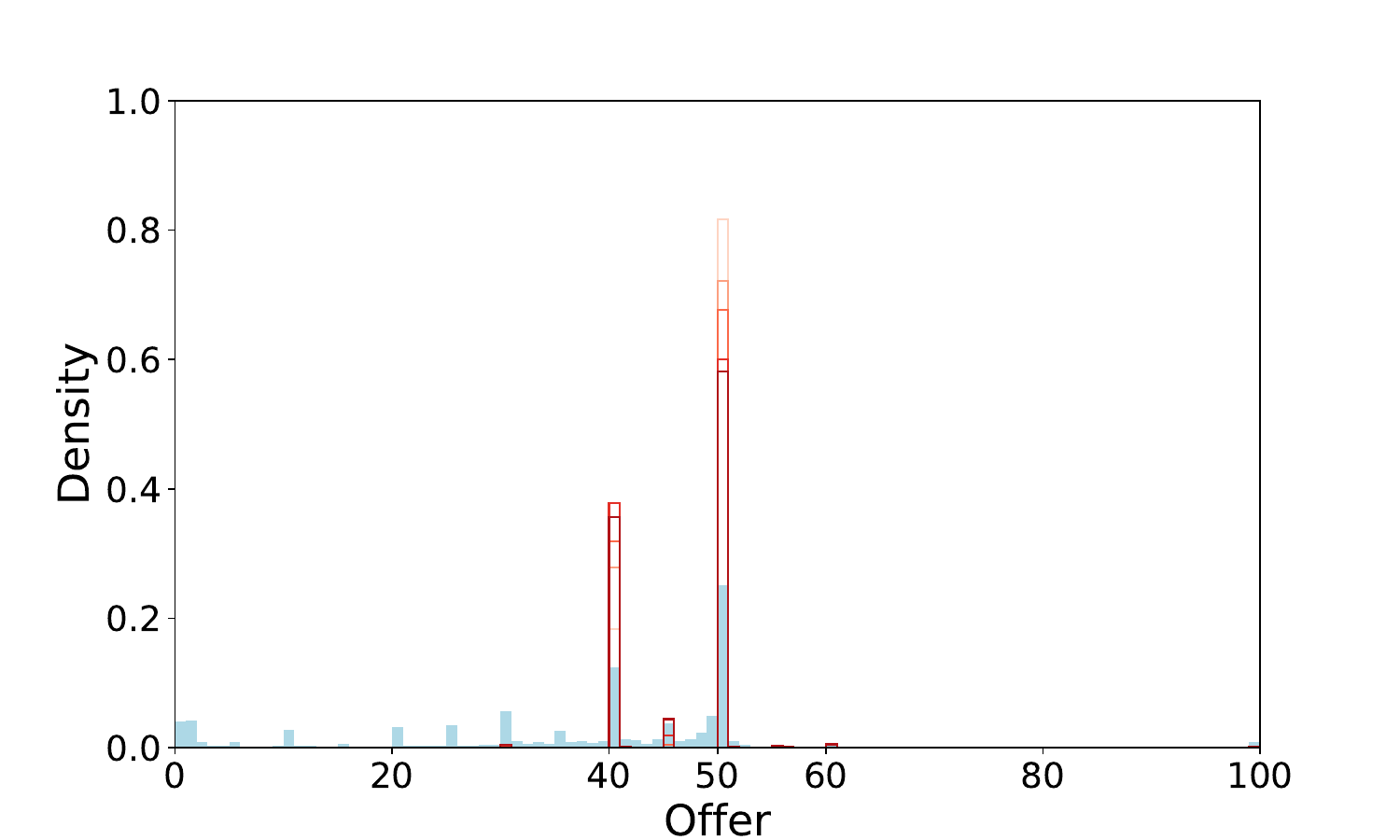}
        \subcaption{Pattern C}\label{subfig:pro-few-4}
    \end{subfigure}
    \begin{subfigure}[t]{0.45\textwidth}  
        \centering	
        \includegraphics[width=\textwidth]{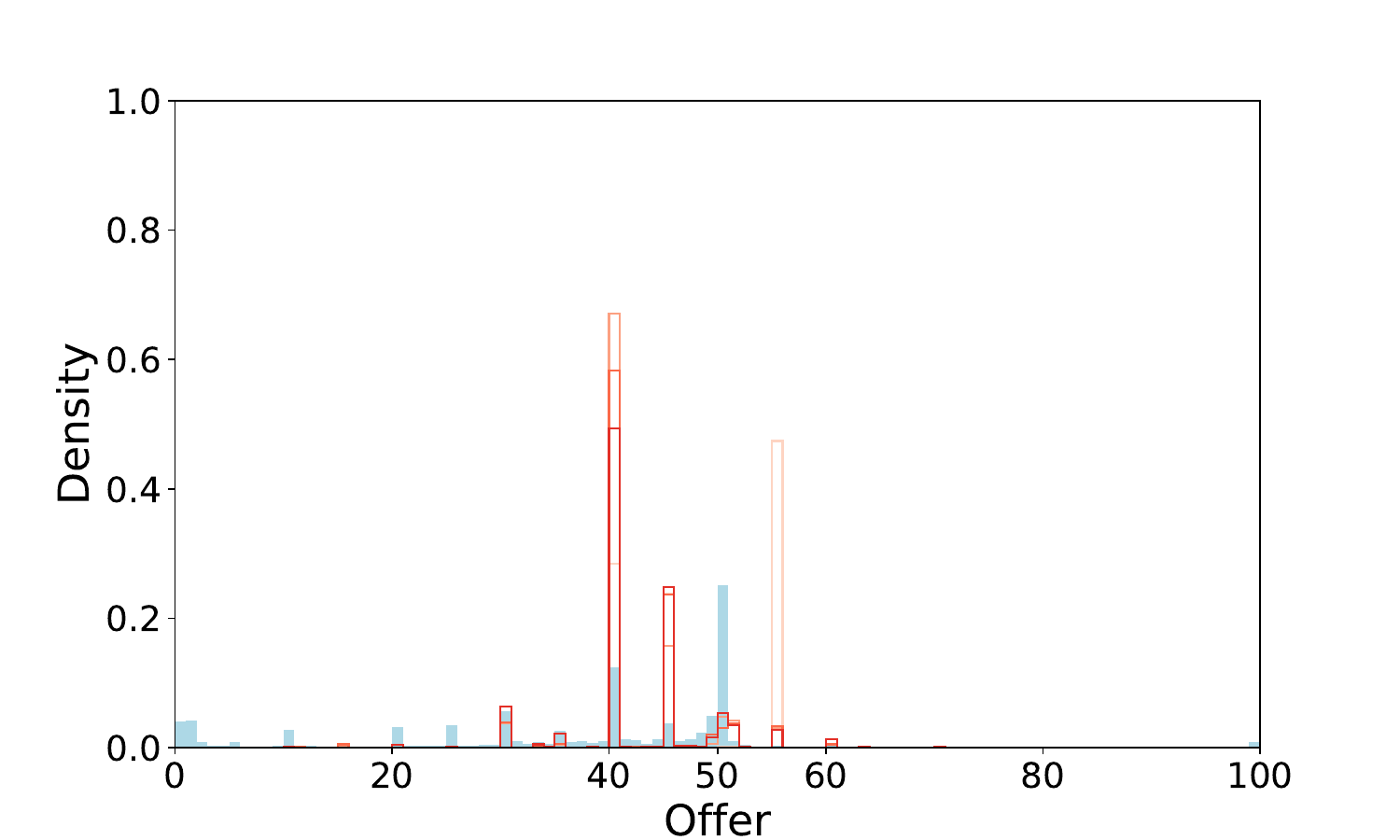}
        \subcaption{Pattern D}\label{subfig:pro-cot-4}
    \end{subfigure}
    \centering
    \caption{Simulation results from the proposer side}
    \label{fig:results-proposer}
\end{figure}

Starting with the proposing player, Figure \ref{fig:results-proposer} presents the results obtained from simulations. The horizontal axis represents the offered value, while the vertical axis represents the normalized count of agents offering each value. The blue histogram visualizes the data from MobLab used in Lin et al. (2020) and remains consistent across all graphs. The simulation results for the five (or four) temperature settings are depicted with histograms outlined in different colors. This figure is composed of four graphs, each one for each pattern setting.

First, Figure~\ref{subfig:pro-zero-3.5} showed similar but a slightly different tendency from our previous study (see Figure~\ref{fig:proposer-pre-result}). In the case of $\text{temperature} = 0$ to $1.5$, the peak appears at $50$ offer, and the higher the temperature setting is, the closer to raw data the distribution is. However, when $\text{temperature} = 2.0$, the distribution did not get closer, showing a higher peak at $50$ offer than when $\text{temperature} = 1.5$. That implies a larger gap with the raw data in comparison with the previous one in Figure~\ref{fig:proposer-pre-result}.

Second, Figure~\ref{subfig:pro-zero-4} showed that in any temperature settings, almost all agents offered $50-50$ division of coins. This is the most frequent offer in the raw data, but the disappearance of other offers in the simulation is a gap with the real economic experiment.

Third, by introducing the other selected prompting methods, other tendencies can be found. When utilizing Few-shot prompting, Figure~\ref{subfig:pro-few-4} shows the most frequent peak at a $50$ offer and the second most frequent peak at a $40$ offer, while there is almost no other offer value. When adapting CoT prompting, in Figure~\ref{subfig:pro-cot-4}, the peak at $50$ offer disappears. Moreover, in the case of $\text{temperature} = 0.5$ to $1.5$, the most frequent peak is at $40$ offer, and the second most frequent peak appears at $45$ offer. However, in the case $\text{temperature} = 0$, the most frequent peak appears at $55$ offer and the second and third most frequent peak is at $40$ and $45$.

\subsection{Responder Side}
\begin{figure}[htbp]
    \centering
    \begin{subfigure}[t]{0.45\textwidth}
        \centering
	\includegraphics[width=\textwidth]{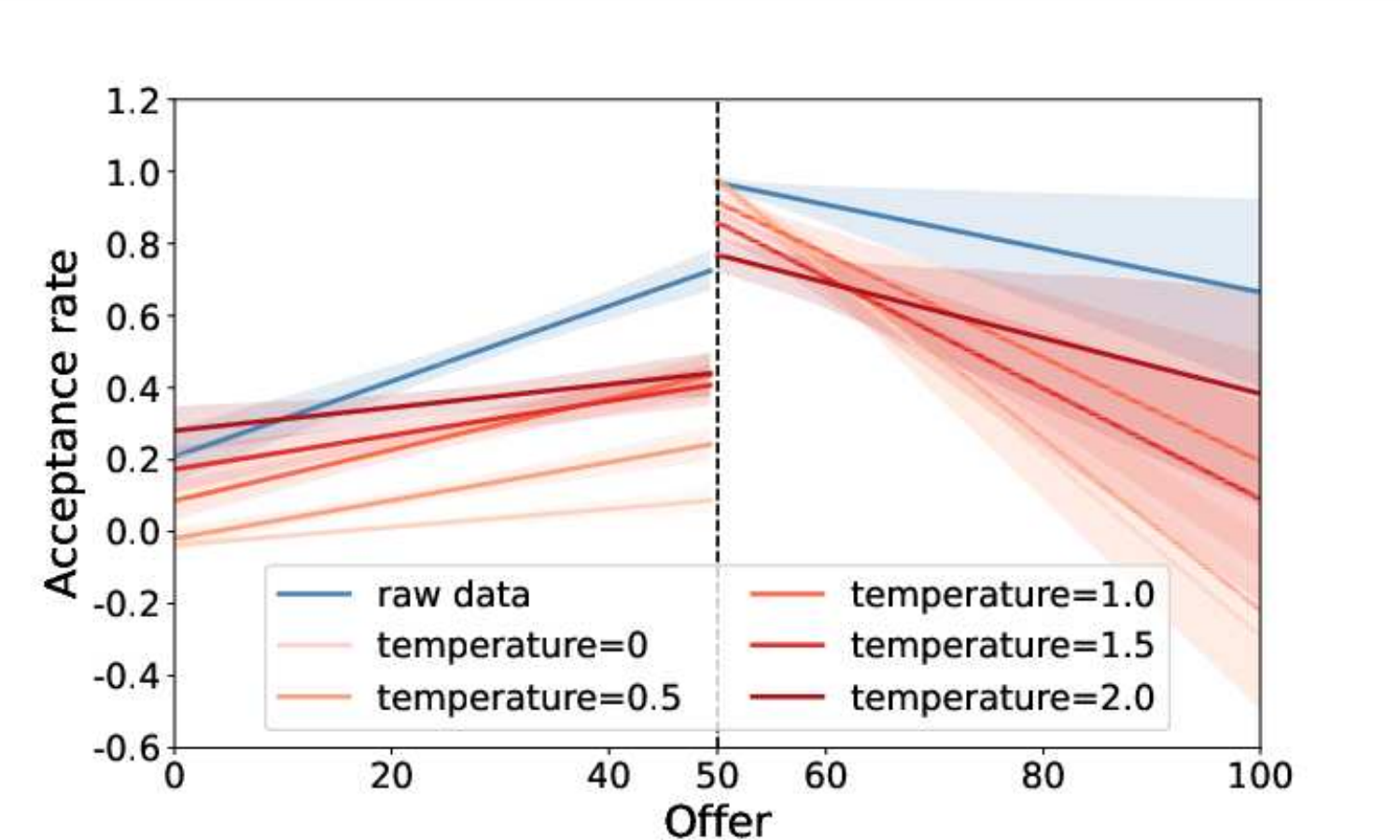}
        \subcaption{Pattern A}\label{subfig:res-zero-3.5}
    \end{subfigure}
    \begin{subfigure}[t]{0.45\textwidth}
	\centering
        \includegraphics[width=\textwidth]{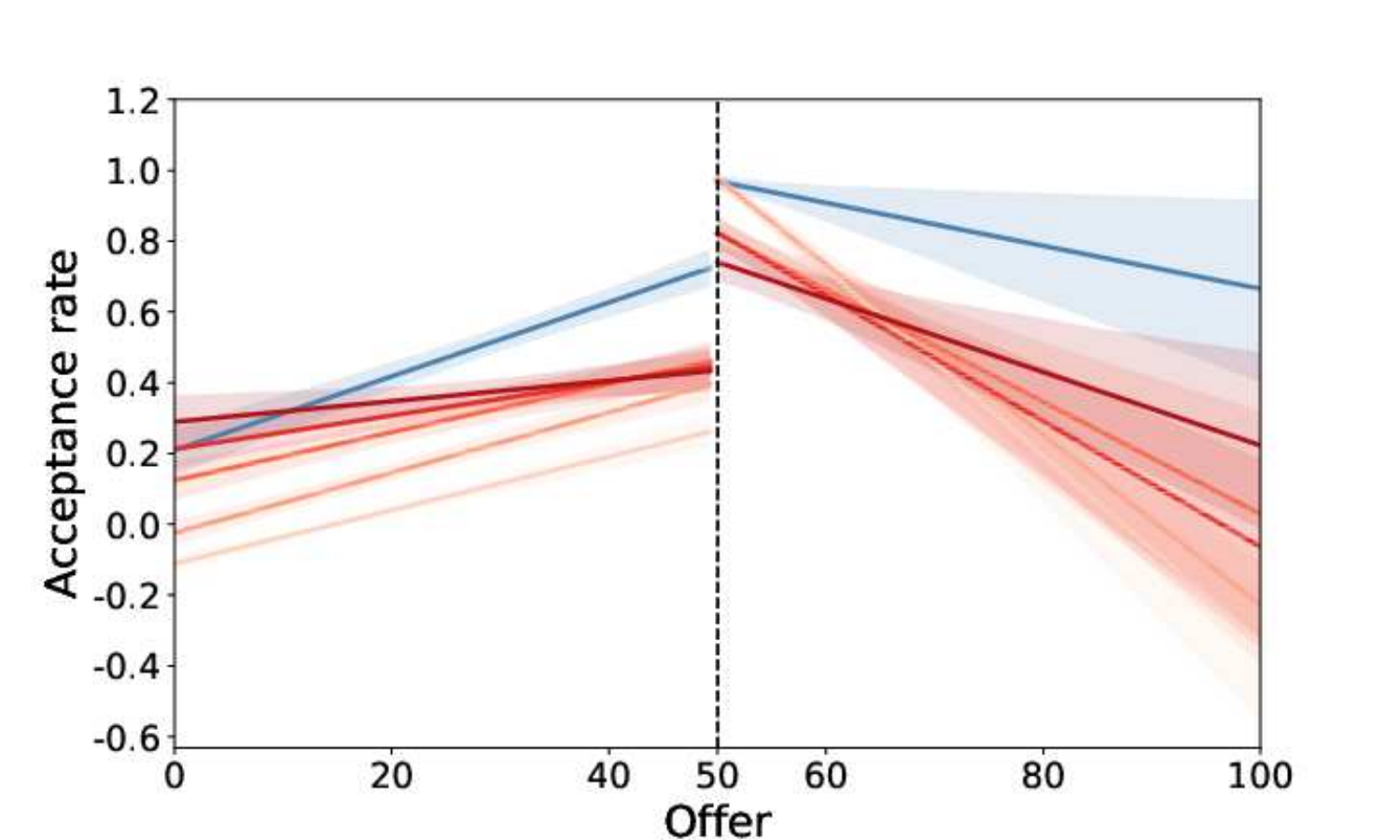}
        \subcaption{Pattern B}\label{subfig:res-zero-4}
    \end{subfigure}
    \begin{subfigure}[t]{0.45\textwidth}    
        \centering
        \includegraphics[width=\textwidth]{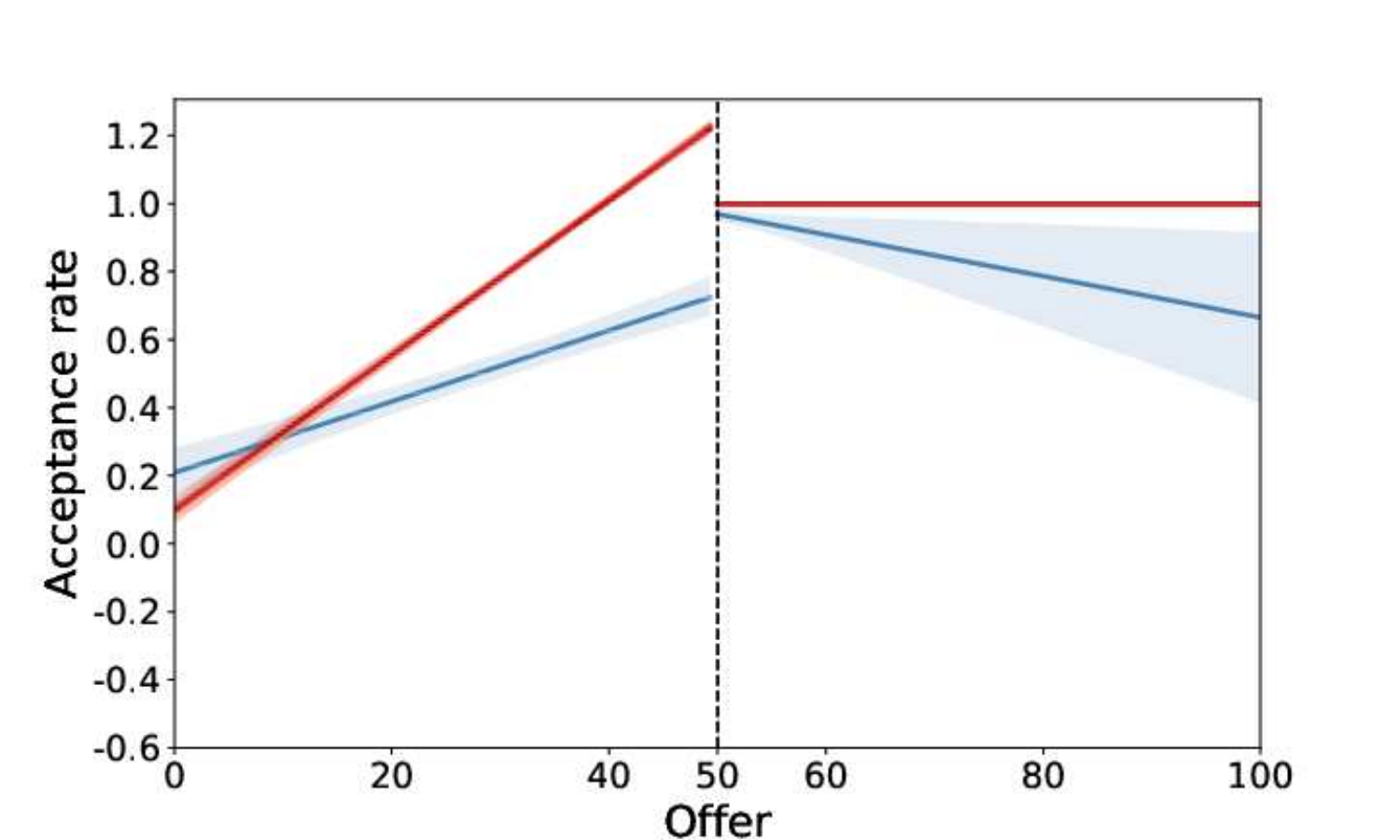}
        \subcaption{Pattern C}\label{subfig:res-few-4}
    \end{subfigure}
    \begin{subfigure}[t]{0.45\textwidth}  
        \centering	
        \includegraphics[width=\textwidth]{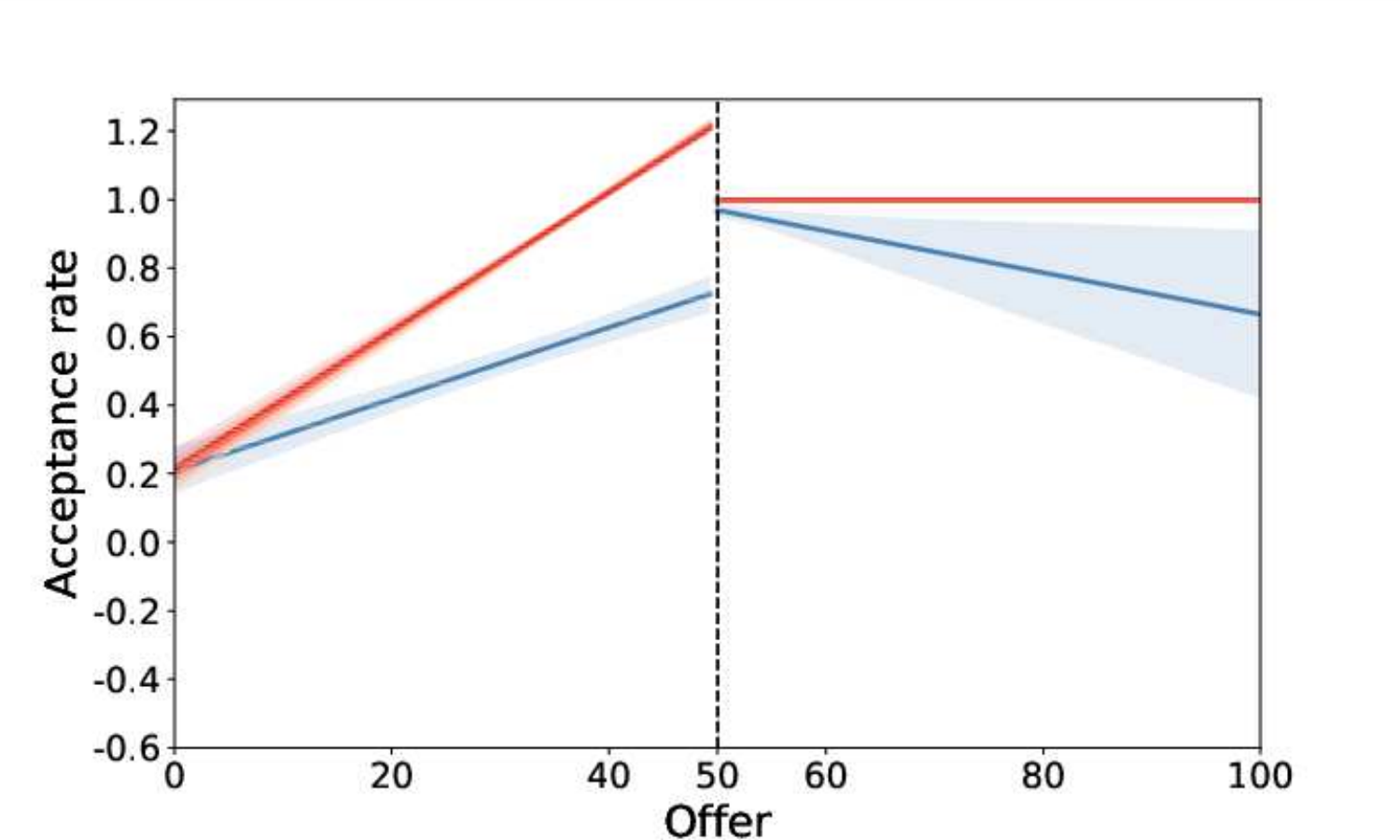}
        \subcaption{Pattern D}\label{subfig:res-cot-4}
    \end{subfigure}
    \centering
    \caption{Simulation results from the responder side}
    \label{fig:results-responder}
\end{figure}

Regarding the responding player, Figure~\ref{fig:results-responder} presents the results from simulations, with one graph for each settings pattern. The horizontal axis of each graph indicates the value offered by the proposer and the vertical axis denotes the acceptance rate for each offer. The blue line represents the data from $1000$ samples out of those used by Lin et al. (2020) and remains consistent across all figures. The simulation results for the five temperature settings are depicted using distinct colored graphs.

Figure~\ref{subfig:res-zero-3.5} and Figure~\ref{subfig:res-zero-4} showed similar tendency. In all patterns of temperature, the higher the offered value is from $0$ to $50$, the higher the acceptance rate is, a discontinuous jump appears at $50$, and the higher the offered value is from $50$ to $100$, the lower the acceptance rate gets. Although this tendency is consistent with the original one, some differences were found. Regarding the left side, the increasing acceptance ratio is smaller than the blue line. At the $50$ offer, the acceptance rate of the left blue line is about $0.7$ and the rate of the other color lines is about $0.4$ or less. Regarding the right side, the acceptance rate of simulations is lower than the row data. In particular, at the $100$ offer, the rates are lower than the one found in $50$. That implies a gap with the actual experimental result.

Next, Figures~\ref{subfig:res-few-4} and ~\ref{subfig:res-cot-4} showed a similar tendency between them but different from the other graphs. In each one, there are slight differences among simulation settings for various values of temperature. On the left side of each graph, simulation lines show a larger increasing acceptance rate against the value of the offer than that of the original one. On the right side, simulation lines are horizontal where the acceptance rate is equal to $1$.

\begin{figure}[htbp]
    \centering
    \begin{subfigure}[t]{0.45\textwidth}
        \centering
	\includegraphics[width=\textwidth]{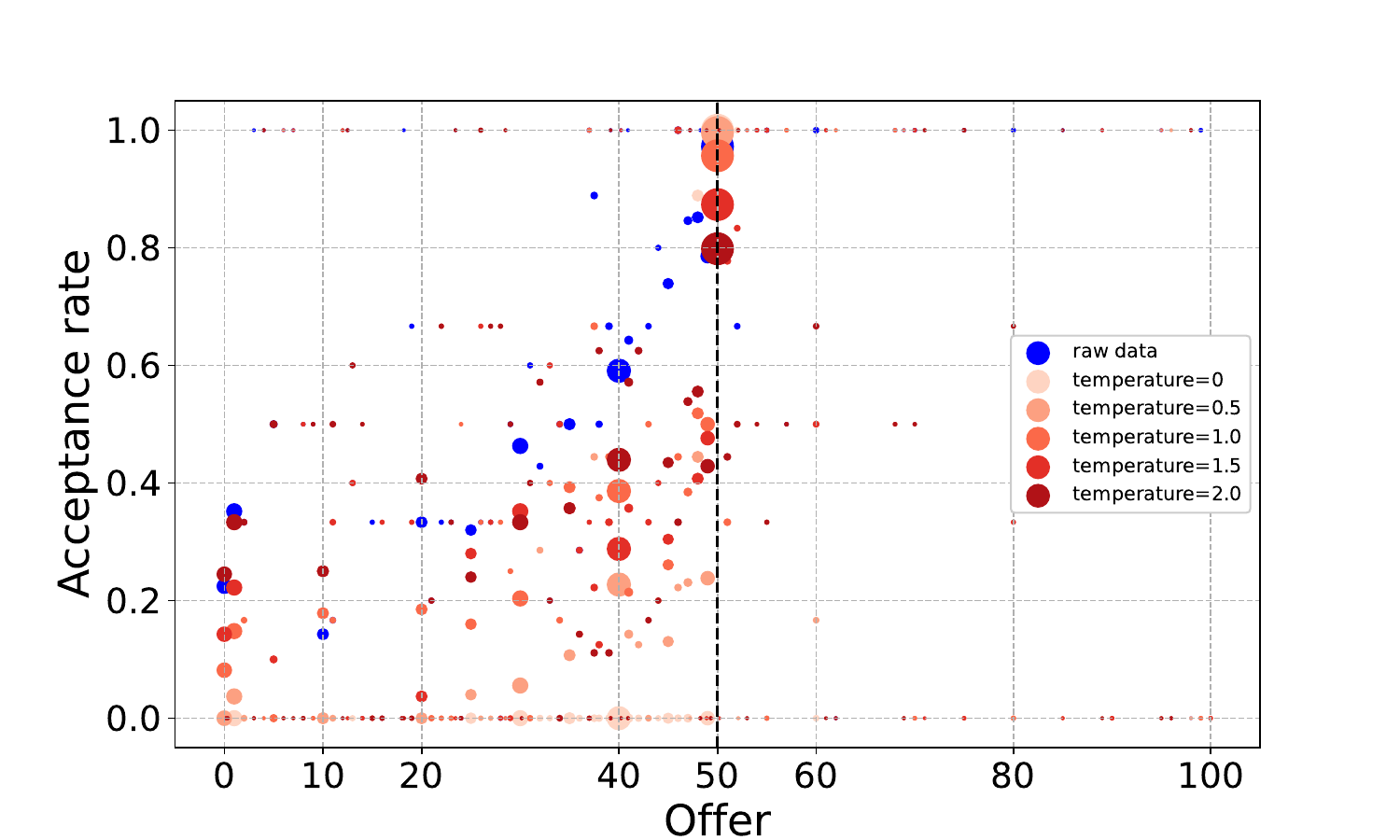}
        \subcaption{Pattern A}\label{subfig:res-zero-3.5-plt}
    \end{subfigure}
    \begin{subfigure}[t]{0.45\textwidth}
	\centering
        \includegraphics[width=\textwidth]{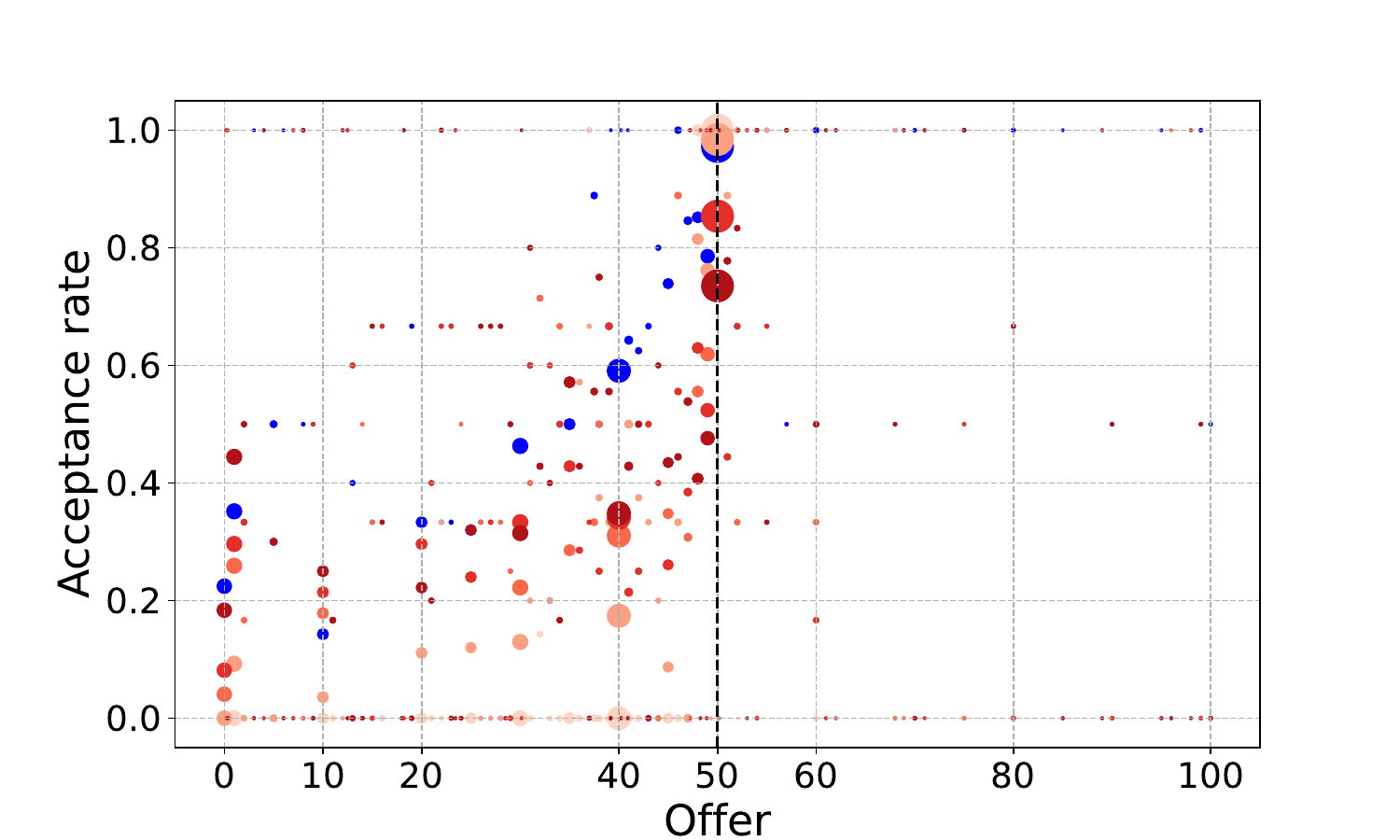}
        \subcaption{Pattern B}\label{subfig:res-zero-4-plt}
    \end{subfigure}
    \begin{subfigure}[t]{0.45\textwidth}    
        \centering
        \includegraphics[width=\textwidth]{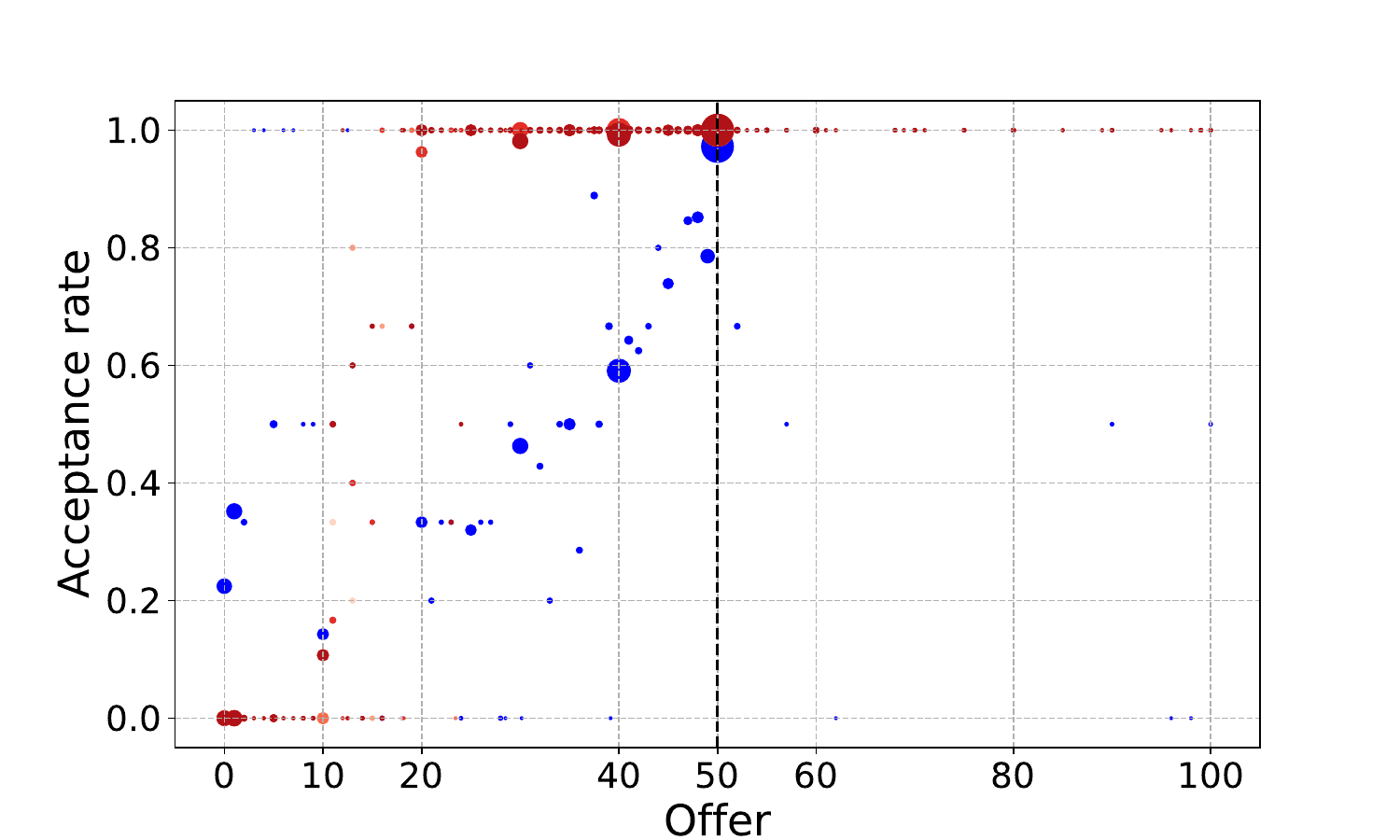}
        \subcaption{Pattern C}\label{subfig:res-few-4-plt}
    \end{subfigure}
    \begin{subfigure}[t]{0.45\textwidth}  
        \centering	
        \includegraphics[width=\textwidth]{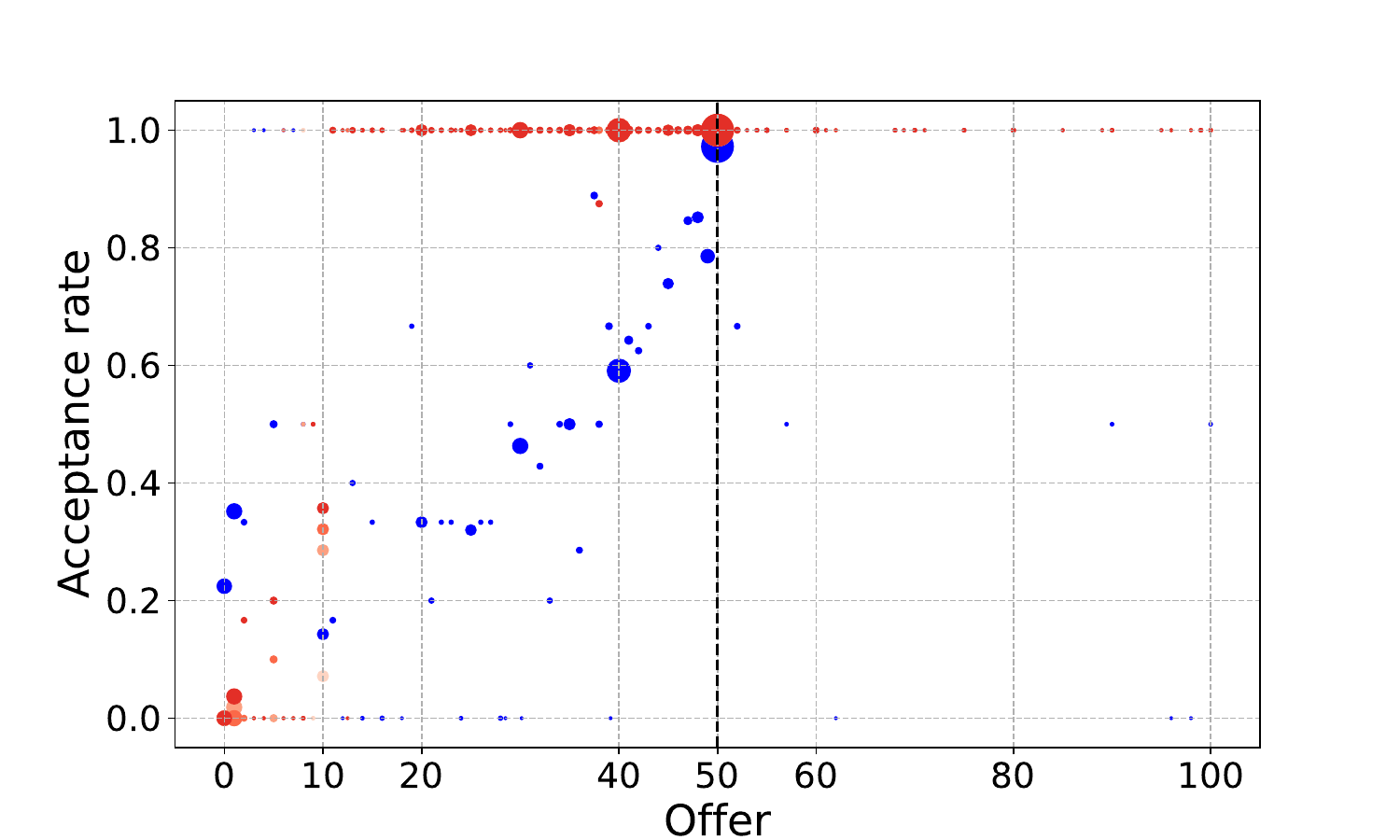}
        \subcaption{Pattern D}\label{subfig:res-cot-4-plt}
    \end{subfigure}
    \centering
    \caption{Bubble Plots of Responder-side Simulation Results}
    \label{fig:results-responder-plt}
\end{figure}

The reason these lines were depicted can be understood by Figure~\ref{fig:results-responder-plt}. This figure presents bubble plots derived from responder-side simulations. The axes and color schemes are consistent with those depicted in Figure~\ref{fig:results-responder}. The center of each bubble represents the acceptance rate corresponding to a specific offer value, while the bubble size reflects the quantity of data points associated with that offer value. These figures reveal that the distributions of data points are skewed. This skewness acounts for the anomalies observed in the regression lines of Figure~\ref{fig:results-responder}, where acceptance rates exhibited extreme values, either exceeding $1.0$ or falling below $0$.

\section{Discussion and Conclusion}\label{sec:discussion}
\subsection{The Effects of Improving Reasoning Ability}
In this research, we conducted four patterns of simulations as a variation of the GPT model and prompt method. Regarding the model of GPT, we used two kinds of models and found that gpt-4-1106-preview is superior to gpt-3.5-turbo-0613 in reasoning ability. As for the prompt method, in decreasing order of reasoning ability, the three methods we used are CoT, Few-shot prompting, and Zero-shot prompting. That is why the four patterns are listed in descending order of the reasoning ability of generative agents as follows: pattern D, pattern C, pattern B, and pattern A.

Considering this, it can be said that the higher the reasoning ability of generative agents is, the closer to the theoretical solution their decision makings become. In the theoretical solution of the ultimatum game, the subgame perfect equilibrium is $(s_1, s_2) = (0, \text{accept})$. Regarding the proposer side, from Figure \ref{subfig:pro-zero-3.5} to Figure~\ref{subfig:pro-cot-4}, the distribution changed toward the offer $0$. Regarding the responder side, by utilizing prompting methods to improve the reasoning ability of generative agents, the fluctuation of acceptance rate in response to an offered value becomes stable and generative agents almost always decide $s_2 = \text{accept}$ when the offered value is larger than $20$ and $s_2 = \text{reject}$ when the offered value is $10$ or less. Therefore, it can be said that by improving the reasoning ability of generative agents, the result of LLMs-driven MAS does not get closer to the actual experimental result, but gets closer to the theoretical solution.

\subsection{Toward the Replication of Human-like Decision Making}
In this research, we tried to reproduce the result of economic experiments found by Lin et al. (2020) through LLMs-driven MAS where the reasoning ability of generative agents was improved. However, as a result, we found that the higher the reasoning ability of generative agents is, the closer not to the experimental result but to the theoretical result the simulation result becomes. In the future, the reasoning ability of LLMs will be further improved following the rapid development of the computer science field. Considering this expectation, this study suggests that more research is needed on getting intelligent generative agents to make a human-like decision. 

Moreover, the comparison of this and our previous research suggests the importance of persona settings. We attribute the difference between Figures \ref{fig:proposer-pre-result} and \ref{fig:responder-pre-result} (a result in our previous work) and Figures~\ref{subfig:pro-zero-3.5} and \ref{subfig:res-zero-3.5} (the result of simulation patterns in this study) to the existence of persona settings. That is because other settings of simulations were completely the same. Comparing these results, the disappearance of the persona setting in this study has negatively worked on the proposer side and positively worked on the responder side for reproducing the result of actual decision making in Lin et al. (2020). This difference suggests that the settings of the persona of generative agents are essential for applying LLMs-driven MAS into economic experiments, which is consistent with previous researches\cite{Horton-2023, Phelps-2023, Han-2023} and needs further investigation. 

Considering this, to get generative agents making human-like decisions how to set the persona of the agents is key. In this research, as an update from Kitadai et al. (2023), we eliminated the step of the persona settings for each agent in the simulation, which generated the discontinuous jump at $50$ offer in Figure \ref{subfig:pro-zero-3.5}. However, when the reasoning ability of generative agents was improved, such elimination caused a larger difference in the results for the proposer side. By setting a proper persona for agents, the tendency of their decision making may get close to that of the actual people. Exploring the general framework to set such persona of generative agents is a future step toward the establishment of a general framework of LLMs-driven MAS to reproduce the result of economic experiments.

\bibliographystyle{ieeetr}
\bibliography{references}  

\end{document}